\documentstyle[aps,multicol,prl,epsfig]{revtex}
\begin{document}
\draft

% Teff01.tex  IKO      10/1/01
% Teff02.tex  AJL      10/5/01
% Teff03.tex  AJL,SRN  10/9/01
% Teff04.tex  everybody 10/11/01

\title{Effective Temperatures of a Driven System Near Jamming}
\author{Ian K. Ono$^1$, Corey S. O'Hern$^{1,2}$, Stephen A.
Langer$^{3}$, Andrea J. Liu$^1$,
and Sidney R. Nagel$^2$}
\address{$^1$~Department of Chemistry and Biochemistry,
University of California, Los Angeles, CA  90095}
\address{$^2$~James Franck Institute, The University of Chicago,
Chicago, IL 60637}
\address{$^{3}$Information Technology Laboratory, NIST,
Gaithersburg, MD 20899}

\date{\today}
\maketitle

\begin{abstract}
      Fluctuations in a model of a sheared, zero-temperature foam are
      studied numerically.  Five different quantities that reduce to
      the true temperature in an equilibrium thermal system are
      calculated.  All five have the same shear-rate dependence, and
      three have the same value.  Near the onset of jamming,
      the relaxation time is the same function of these three
      temperatures in the sheared system as of the true temperature in
      an unsheared system.  These results imply that statistical
      mechanics is useful for the system and provide strong support for
      the concept of jamming.
\end{abstract}

\begin{multicols}{2}
\narrowtext

Statistical mechanics describes the connection between microscopic
properties and collective many-body properties in systems in thermal
equilibrium.  There is no equivalent formalism for driven, athermal
systems.  Nonetheless, recent phenomenological
approaches\cite{sollich} that assume thermal behavior are
surprisingly successful in describing driven glassy materials such as
sheared foam.  Foam is a dense packing of bubbles in a small amount of
liquid, and is athermal because the thermal energy is much smaller
than the typical energy barrier for bubbles to change their relative
positions\cite{book}.  As a result, quiescent foam is
{\it jammed}\cite{book}; it is disordered and has a yield stress.  If
foam is steadily sheared, however, it is pushed over energy
barriers and flows as different bubble packings are explored. However,
it is unclear if this degree of ergodicity is enough to lead to
thermal behavior.

\begin{figure}
\epsfxsize=3.2in \epsfbox{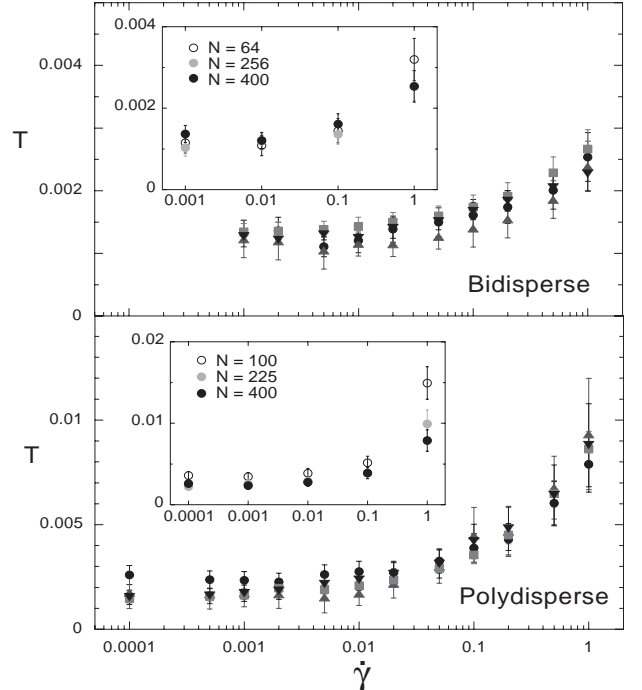}
\caption{Four effective temperatures, $T_{p}$ (circles), $T_{xy}$
(upward triangles), $T_{E}$ (squares) and $T_{f}$ (downward triangles)
calculated as a function of shear
rate.
$T_{f}$ has been rescaled by 4 and 2.7 for the polydisperse and
bidisperse systems, respectively, to collapse on the others.  Insets: $T_{p}$
vs. $\dot \gamma$ for different system sizes.}
\label{teff}
\end{figure}

In this Letter, we test the assumption that a sheared foam can be
modeled as a thermal system with a temperature that depends on
shear rate.  We conduct numerical simulations of a simple
model of sheared foam and measure five quantities that all reduce to
the true temperature in a thermal system.  Although these quantities
must all have the same value in an equilibrium thermal system, there
is no guarantee that they should be the same in the steadily-sheared
model foam.  Remarkably, three of the
effective temperatures are the same and all five have the same
shear-rate dependence.  Our results for four of the effective temperatures
are shown as
a function of shear rate in Fig.~\ref{teff} for two different size
distributions of the bubbles.  In a companion
paper, Berthier and Barrat reach similar conclusions for a sheared
thermal system\cite{berthier1}.  These results suggest that statistical
mechanics is indeed useful for describing driven jamming systems.

Our bubble dynamics (BD) simulations are carried out in two 
dimensions on Durian's model of
foam\cite{doug}. The bubbles are circles with diameters assigned from
one of two different diameter distributions.  The first (polydisperse)
distribution is flat from 0.2 to 1.8 times the average bubble diameter
and is zero otherwise.  The second (bidisperse) distribution consists
of equal numbers of small and large bubbles of diameter ratio 1.4.
Pairs of bubbles only interact, via a repulsive spring, when they
overlap;
this approximates the energy cost of bubble deformation\cite{doug}.
There is also a frictional force proportional to the velocity 
difference between a bubble and
the average flow at its position.  The system is fully periodic, with
flow in the $\hat x$ direction and a shear gradient in the $\hat y$
direction imposed using the Lees-Edwards boundary
condition\cite{allen}.
%%%%%%%%%%%%%%%%%
%%%%%%%%%%%%%%%%%

We scale lengths by the average diameter $d$, energies by $k
d^{2}$ where $k$ is the spring constant, and time scales by
$\tau_{0}=b/k$, where $b$ is the friction coefficient.  Thus, the
dimensionless shear rate $\dot \gamma$ is the Deborah number.
Unless otherwise specified, the systems contain
$N=400$ bubbles at an area fraction of $\phi=0.9$ (well above random
close-packing at 0.84).  Averages are typically taken over time
snapshots separated by 0.1 in strain over a total strain of 10 for 100
(polydisperse) or 2 (bidisperse) different initial configurations.
%%%%%%%%%%%%%%%%%
\begin{figure}
\epsfxsize=3.2in \epsfbox{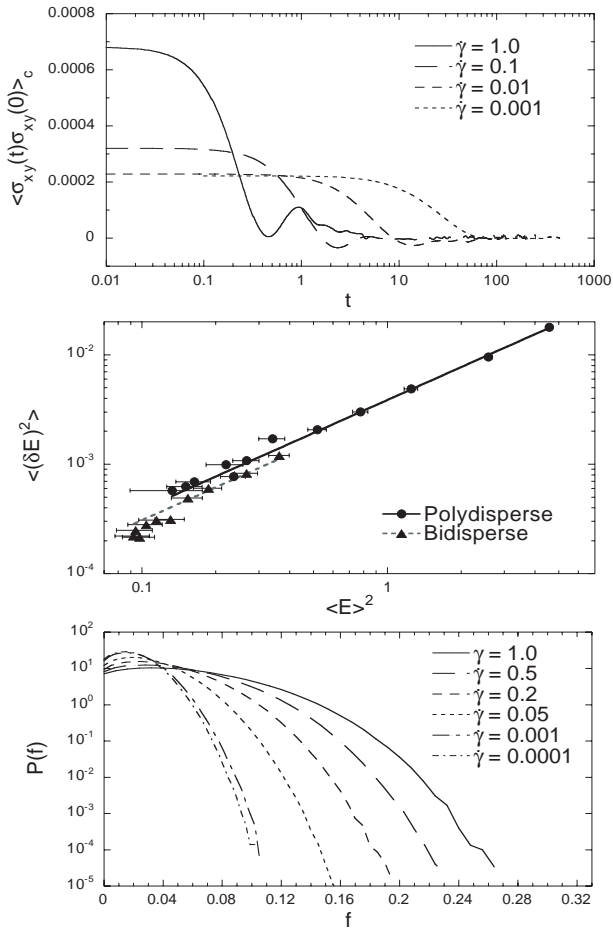}
\caption{(a) Shear stress autocorrelation functions at different shear
rates for the polydisperse system (see Def.~2).  (b) Variance
of energy fluctuations plotted against average energy squared (Def.~3).
The lines are fits to $\langle (\delta E)^{2} \rangle = m \langle E
\rangle^{2}$, where $m=3.9 \times 10^{-3}$ (polydisperse).  (c)
Force distributions $P(f)$ at different shear rates for the
polydisperse system.  The tail of
$P(f)$ is fitted to a Gaussian centered at zero to obtain the
effective temperature $T_{f}$ (Def.~4).}
\label{support}
\end{figure}
%%%%%%%%%%%%%%%%%

Three of the 5 calculated effective temperatures are based on linear
response relations or fluctuation-dissipation relations.  Similar
definitions have been used to characterize non-equilibrium systems in
the contexts of weak turbulence\cite{hohenberg}, aging of glassy
systems\cite{leticia}, granular packings\cite{nowak,makse} and sheared
aging systems\cite{book,berthier2}; such effective temperatures can 
control the direction of heat
flow and thus play the role of temperature in the thermodynamical
sense in certain nonequilibrium systems with small energy
flows\cite{ckp}.

{\it Definition 1: pressure fluctuations.}  In an equilibrium system
at fixed $N$, $T$ and area $A$,
the variance of the pressure is given
by\cite{allen}
\begin{equation}
      \langle p \rangle + {\langle x \rangle \over A} - \beta_{T}^{-1}
={A \over T} \langle
(\delta p^{2})\rangle
\label{equilpdef}
\end{equation}
where $\beta_{T}^{-1} \equiv - A (\partial
\langle p \rangle / \partial A)_{T}$ is the inverse isothermal
compressibility, $A$ is the area of the system, $p$ is the pressure, $x$ is the
hyper-virial\cite{allen}, and the Boltzmann constant is unity.
In our driven, athermal system we thus define
\begin{equation}
     T_{p}={A  \langle(\delta p^{2})\rangle \over \langle p \rangle +
     {\langle x \rangle \over A} - \beta_{T}^{-1}},
     \label{tpdef}
\end{equation}
so that $T_{p}$ reduces
to the true temperature in an
equilibrium thermal system.
To measure the compressibility, we perturb the system area
$A$ and measure the resulting value of $\langle p \rangle$ during
shear to calculate
the derivative. In simulations of a quiescent system in thermal equilibrium
with the same potential and polydispersity, we find that with comparable
statistics, Eq.~\ref{equilpdef} yields results within 5\% of the
simulation temperature.

{\it Definition 2: shear stress fluctuations.} The viscosity of an
equilibrium system is related to the integral over the shear stress
autocorrelation function\cite{allen}:
\begin{equation}
      \eta={A \over T} \int_{0}^{\infty} dt \langle \sigma_{xy}(t)
      \sigma_{xy}(0) \rangle_{c}
      \label{equiletadef}
\end{equation}
Again, we use Eq.~\ref{equiletadef} to define $T_{xy}$,
calculating the steady-state shear viscosity from
$\eta=\langle \sigma_{xy} \rangle/\dot \gamma$.  The stress 
autocorrelation function is
shown for several different shear rates in Fig.~\ref{support}(a).
With decreasing $\dot \gamma$, the
correlation time increases approximately linearly, while the
variance $\langle (\delta \sigma_{xy})^2
\rangle$ decreases and then saturates.

{\it Definition 3: energy fluctuations.} The constant-volume heat
capacity of an equilibrium system is related to energy fluctuations:
\begin{equation}
      {d \langle E \rangle \over dT}={\langle (\delta E)^{2} \rangle
      \over T^{2}}.
      \label{equilcvdef}
\end{equation}
This can be rearranged and integrated on both sides to provide a
definition of $T$\cite{nowak}.  To calculate $T_{E}$ we must extract
the relation between the variance of the energy fluctuations, $\langle
(\delta E)^{2} \rangle$, and the average energy, $\langle E \rangle$.
The results are shown in Fig.~\ref{support}(b).  We find that the
variance scales as the square of the average energy, as shown by the
line-fits to the data.  The fits imply that $\langle E \rangle
\propto T_{E}$, as can be seen by substitution into
Eq.~\ref{equilcvdef}.
In fact, we find $\langle E \rangle = 0.64 N T_{E}$ for the
polydisperse system and $\langle E \rangle = 0.56 N T_{E}$ for the
bidisperse system, where $N$ is the number of bubbles in our
two-dimensional system.  These results resemble equipartition, except
that the coefficient of $NT$ is not unity.
Our potential is a harmonic repulsion with finite range, so 
equipartition is not
exact.

{\it Definition 4:  force distribution.}
The temperature can be extracted from the tail
of the distribution of interparticle normal forces $P(f)$\cite{ohern}.  The
force distribution is directly related to
the pair correlation function $g(r) \equiv \exp(-\beta u(r)) y(r)$ in a system
with a pair
potential $u(r)$.  For $r$ sufficiently
small, the exponential term is a much stronger function of $r$ (and
hence of $f$) than $y(r)$.  As a result, the tail of $P(f)$
depends on the interparticle potential and the temperature:
\begin{equation}
      P(f) \approx \exp(-f^{2}/2T).
      \label{pftail}
\end{equation}
We fit the tail to extract $T_{f}$.  Eq.~\ref{pftail} applies only to
a monodisperse system.  For a bidisperse system, we measure the
distribution for small particles interacting with small particles
only.  For the polydisperse case, however, we use the force
distribution for all particles (Fig.~\ref{support}(c)).  From MD
simulations of an equilibrium system with the same potential, we have
verified that this assumption should lead to no more than a 30\% error
in the computed $T_{f}$.

%%%%%%%%%%%%%%%%%
\begin{figure}
\epsfxsize=3.2in \epsfbox{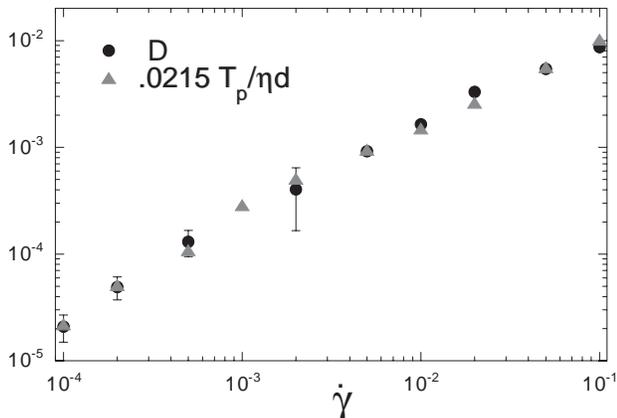} \caption{The Stokes-Einstein relation.
The circles represent the diffusion constant $D$,
measured by integrating the velocity autocorrelation function.  The
triangles represent $T_{p}/C \eta d$, where $C$ is a constant chosen
to obtain the best fit with $D$.}
\label{SEcheck}
\end{figure}
%%%%%%%%%%%%%%%%%

{\it Definition 5:  Stokes-Einstein relation.}
In equilibrium, the diffusion constant $D$ satisfies
\begin{equation}
      D={T \over C \eta d},
      \label{SE}
\end{equation}
where $C$ depends on system size in two
dimensions.  Thus, Eq.~\ref{SE} defines a
temperature $T_{D}$ up to an unknown constant.   Here we use fixed boundary
conditions in the $y$-direction\cite{steve}.  We measure $D$
in the $y$-direction in two different ways and find good agreement: we
integrate the velocity autocorrelation function, and we measure the
displacement distribution as a function of time from an initial
starting position.   It is difficult to extract $T_{D}$
because $D$ and $\eta$ vary by several orders of
magnitude over the range of $\dot\gamma$ studied, while their product varies
by less than an order of magnitude and has a lot of scatter and large
error bars.   This definition is
therefore most useful as a consistency check:  we use $T_{p}$ from
Def. 1 and vary $C$ to obtain the best agreement between the
left and right sides of Eq.~\ref{SE}.  Fig.~\ref{SEcheck}
shows that the Stokes-Einstein relation is indeed obeyed.

Our results for the other four effective temperatures (Defs.~1-4) are
plotted in Fig.~\ref{teff}.  The insets show that the temperatures do
not depend on system size for $N$ sufficiently large.  All the
temperatures have the same shear rate dependence over 4 decades of
$\dot\gamma$.  Since we do not have a first-principles calculation of
$C$, the magnitude of $T_{D}$ is not known.  However, $T_{f}$ is different
in magnitude (but not in $\dot\gamma$-dependence) from the remaining
three.  One possible reason for the discrepancy is that $T_{f}$
measures the properties of the {\it tail} of $P(f)$, while $T_{p}$,
$T_{xy}$ and $T_{E}$ measure fluctuations around average quantities.
This suggests that the underlying probability distribution may not be
a Boltzmann distribution.

Fig.~\ref{teff} suggests that the effective temperatures approach
nonzero constants $T_{\rm eff}^{0}$ in the limit $\dot \gamma \rightarrow
0$.  In fact, this is expected from their definitions.  As long as
bubbles overlap as $\dot \gamma \rightarrow 0$, the force
distribution $P(f)$ is nonzero for $f>0$, and yields
$T_{f}^{0}>0$.  The other temperatures should also be nonzero; in the
zero shear rate limit, the angular brackets indicate configurational
averages rather than time averages.  This suggests an interpretation
of the limiting value; $T_{\rm eff}^{0}$ should correspond to the 
glass transition
temperature $T_{g}$.  By shearing the system and calculating averages
over times long compared to the relaxation time (which scales as
$1/\dot \gamma$), we are demanding that the system is ergodic.
Therefore, the only temperatures accessible to us are above $T_{g}$.

\begin{figure}
\epsfxsize=3.2in \epsfbox{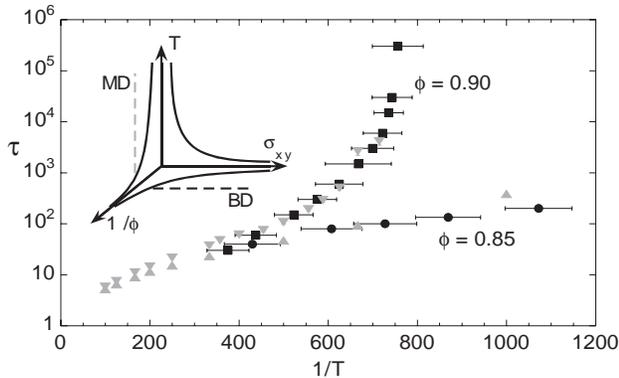}
\caption{An
Arrhenius plot of the
relaxation time as a function of temperature from MD (gray symbols) and BD
(black symbols) simulations at two different packing fractions, $\phi=0.85$ and
$\phi=0.90$.  Inset:
the jamming phase diagram.  Our bubble dynamics simulations
follow the trajectory marked BD, while our molecular dynamics
simulations follow the trajectory marked MD.}
\label{mdvsbd}
\end{figure}

We have checked the interpretation of $T_{\rm eff}^{0}$ as $T_{g}$ by
conducting equilibrium molecular dynamics (MD) simulations on a system
with the same interaction potential, packing fraction, and size distribution
(bidisperse) as in our bubble dynamics (BD) runs.  It has been
proposed that jamming systems such as this one can be described by a
phase diagram\cite{lucid}, sketched in the inset to Fig.~\ref{mdvsbd}.
This diagram shows that jamming occurs as $T$ is lowered, the packing
fraction $\phi$ is raised, or the applied shear stress $\sigma_{xy}$ is
lowered, and has been shown to be a useful way to represent
experimental data\cite{trappe}.  The BD results in Fig.~\ref{teff}
correspond to the trajectory marked ``BD'' in the inset to
Fig.~\ref{mdvsbd}.  In the MD simulations, we have removed the
frictional term from the equations of motion and added the inertial
term and true temperature so as to approach jamming along the
trajectory marked ``MD.''  In
Fig.~\ref{mdvsbd}, we show the results for the relaxation time $\tau$
as a function of $T$ from MD and BD, at two different packing
fractions, $\phi=0.85$ and $\phi=0.90$.  The lower packing fraction is just
above random close-packing.  In MD, we measure $\tau$ from the decay
of the intermediate scattering function\cite{mdglass}.  In BD, we plot
$\tau=c/\dot\gamma$, where $c$ is a constant chosen to best fit the MD
data\cite{remark}, as a function of $T_{E}$.

Fig.~\ref{mdvsbd} shows that the dynamics
are the {\it same} for a system approaching jamming by two different
trajectories: decreasing $T$ and decreasing $\dot \gamma$ (or
equivalently, decreasing $\langle \sigma_{xy} \rangle$).  Thus,
$T_{g}$ from MD at zero applied shear is the same as $T_{\rm eff}^{0}$
from BD. A similar result was found previously for a sheared thermal
Lennard-Jones mixture\cite{berthier2}.  Fig.~\ref{mdvsbd} also shows
that the {\it functional form} of the slowing down of the dynamics is
the same along both trajectories for low $T$.  This is consistent with
previous work showing that the dynamics of the sheared model foam
can be described by a Vogel-Fulcher form\cite{steve}.  An attempt to
collapse the data for the two different $\phi$ using an Angell
fragility plot\cite{angell} shows that the fragility depends on 
packing fraction.

%%%%%%%%%%%%%%%%%
%%%%%%%%%%%%%%%%%

The results
in Fig.~\ref{mdvsbd} provide confirmation of the idea underlying the
jamming phase diagram, namely, that
the jammed region controls the behavior nearby so the dynamics should
not depend on the direction along which the jammed region is
approached\cite{lucid,book}.  It also suggests that one can collapse the
$\sigma-$axis onto the $T-$axis using $T_{\rm eff}$, and
that a system will jam once fluctuations, whether thermal or
shear-induced, are sufficiently small.

Our main finding that 5 different effective temperatures have the same
$\dot \gamma$ dependence raises the need for a criterion for when the
concept of effective temperature might be useful.  We suggest such a
criterion based on the idea underlying the fluctuation-dissipation
relation.  An analogous concept applies to a steady-state driven
system, because the average power supplied to the system must be
balanced by the average power dissipated.  The power can be dissipated
in two ways--by the average flow and by fluctuations around the
average flow.  We speculate that the concept of effective temperature
is useful if nearly all the power supplied by the driving force
is dissipated by fluctuations.  In the model
studied here, all of the power is, by construction, dissipated by
fluctuations--the frictional force is proportional to the difference
between the velocity of a bubble and the average shear.  In
systems in the stick-slip regime near jamming, fluctuations typically are
large compared to the average flow\cite{book}.  This suggests that 
the concept of
effective temperature should be useful for {\it any} system near the onset of
jamming.

We are grateful to B. C. H. Ng for carrying out some of the
runs.  We thank D. J. Durian and C. M. Marques
for instructive discussions.  This work was supported by
NSF-DMR-0087349 (IKO,CSOH,AJL) and NSF-DMR-9722646 (CSOH,SRN).

\end{multicols}

\begin{references}
\bibitem{remark} The time scales in the BD and MD simulations have 
different units
because there is a friction coefficient in BD and a particle mass in
MD, so we are free to choose a scale factor for $\tau$.
\end{references}
\end{document}